\documentclass[prb,preprint,showpacs,floatfix]{revtex4}
\usepackage{graphicx}
\usepackage{epsf}
\usepackage{ifthen}
\usepackage{amsmath}
\usepackage{amssymb}
\usepackage{subfigure}
\usepackage{psfrag}
\usepackage{float}
\usepackage{subeqnarray}
\usepackage{color}
\usepackage{booktabs}
\begin{document}

\title{Optical Properties of GaS-Ca(OH)$_2$ bilayer heterostructure}

\author{E. Torun}
\email{engin.torun@uantwerpen.be}
\affiliation{Department of Physics, University of Antwerp, Groenenborgerlaan 171, 2020 Antwerp, Belgium}

\author{H. Sahin}
\affiliation{Department of Physics, University of Antwerp, Groenenborgerlaan 171, 2020 Antwerp, Belgium}

\author{F. M. Peeters}
\affiliation{Department of Physics, University of Antwerp, Groenenborgerlaan 171, 2020 Antwerp, Belgium}

\date{\today}


\begin{abstract}
Finding novel atomically-thin heterostructures and understanding their characteristic properties are critical for developing better nanoscale optoelectronic devices. In this study, we investigate the electronic and optical properties of GaS-Ca(OH)$_2$ heterostructure using first-principle calculations. The band gap of the GaS-Ca(OH)$_2$ heterostructure is significantly reduced when compared with those of the isolated constituent layers. Our calculations show that the GaS-Ca(OH)$_2$ heterostructure is a type-II heterojunction which can be used to separate photoinduced charge carriers where electrons are localized in GaS and holes in the Ca(OH)$_2$ layer. This leads to spatially indirect excitons which are important for solar energy and optoelectronic applications due to their long lifetime.  By solving the Bethe-Salpeter equation on top of single shot GW calculation (G$_0$W$_0$) the dielectric function and optical oscillator strength of the constituent monolayers and the heterostructure are obtained. The oscillator strength of the optical transition for GaS monolayer is an order of magnitude larger than Ca(OH)$_2$ monolayer. We also found that the calculated optical spectra of different stacking types of the heterostructure show dissimilarities, although their electronic structures are rather similar. This prediction can be used to determine the stacking type of  ultra-thin heterostructures.
\end{abstract}

\maketitle

\section{Introduction}
The successful synthesis of graphene was a milestone in 
condensed matter physics and materials science. 
\cite{graphene1,graphene2,graphene3} Due to this remarkable achievement, a 
new field of quasi-two-dimensional (2D) materials has emerged which 
changed the perspective of materials research. Since then, the attention of 
material science and condensed matter physics has been widened towards new 
single layer structures such as silicene \cite{guzman,seymur}, germanene 
\cite{seymur,houssa,sahin}, transition-metal dichalcogenides (TMDs) 
\cite{neto,mak,splendiani,wang},  alkaline-earth-metal hydroxide (AEMHs) 
\cite{port} and post-transition metal chalcogenides (PTMCs).\cite{ma,ptmcs,ptmcs2} 

These individual monolayers posses various significant electronic and 
optical properties which make them promising candidates for the next 
generation of nanoscale devices.\cite{ataca,miro,chhowalla} For instance, monolayer TMDs are direct band gap semiconductors 
unlike their bulk counterparts, \cite{mak,splendiani,wang} and exhibit large exciton binding energies in the order of 0.1-1.0 eV which results in  
 exciton resonances at room temperature. \cite{he,klots,chernikov,ugeda} In addition, 
having strong coupling between the spin and the valley degrees of freedom 
opens up the possibility of valleytronic devices. \cite{heinz,sallen,xu}     

In spite of the large amount of research on graphene-like structures and 
ultra-thin TMDs, studies on single layer AEMHs (i.e.~Ca(OH)$_2$ and 
Ni(OH)$_2$) and PTMCs (i.e.~GaS and GaSe) are sparse and have only very recently 
started. In the recent study of Aierken \textit{et al.} it was shown that 
Ca(OH)$_2$ can be isolated in monolayer form and it is a 
direct band gap semiconductor independent of the number of layers. \cite{port} 
Similar to Ca(OH)$_2$, monolayer GaS and GaSe have been synthesized recently and it has been shown that both monolayers 
are indirect gap semiconductors and they are suitable candidates for use in 
field-effect transistors and nanophotonic devices. \cite{ptmcs,late1,late2,hu}

Another important aspect of the mentioned layered structures is 
their usage as building blocks for novel multi-layer heterostructures.  Recently, 
a new field of research in materials science has emerged that deals with the 
stacking of two or more different monolayers on top of each other, namely van 
der Waals (vdW) heterostructures.\cite{geim} This widens considerably the 
diversity of possibilities for new functionalities and increases the range of 
different electronic and opto-electronic applications. For instance, it has 
been shown that a p-n junction based on the hBN-WSe$_2$ heterostructure
exhibit tunable electroluminescence \cite{ross}. In addition, TMD 
heterostructures show a tunable photovoltaic effect \cite{furchi} and can be 
used as tunnel diodes and transistors \cite{roy}. Long-lived interlayer 
excitons were observed by photoluminescence excitation spectroscopy in 
MoSe$_2$-WSe$_2$ heterostructure where electrons and holes are localized in 
different layers. \cite{rivera} It has been shown that spatially indirect 
excitons can also be found in MoS$_2$-WSe$_2$ heterostructures which is a type II 
heterojuction. \cite{fang}  Very recently, Calman \textit{et al.} showed that the excitons in MoS$_2$ and 
hexagonal boron nitride (hBN) vdW heterostructure can be controlled by a gate 
voltage, temperature, and the intensity and the helicity of the optical 
excitation. \cite{calman} 

So far, the building blocks of these multi-layer heterostructures are 
usually graphene (and graphene-like monolayers) or 
TMD monolayers. Whereas, in this work we propose a new kind of 
heterostructure whose building blocks are GaS and Ca(OH)$_2$ which are 
relatively new 2D crystals and they are members of PTMCs and AEMHs, 
respectively. We show that the GaS and Ca(OH)$_2$ monolayers have similar 
lattice parameters and they are wide band gap semiconductors. When they are 
stacked on top of each other they form a type II heterojuction which has a 
smaller band gap than the constituent layers. The calculated optical oscillator 
strength of the GaS monolayer is $\sim$10 times larger than the 
one for  monolayer Ca(OH)$_2$. 

This paper is organized as follows:
We first provide the computational methodology in Sec.~II. Then, 
we investigate the geometric, electronic and optical properties 
of the isolated GaS and Ca(OH)$_2$ monolayers in Sec. III and the GaS-Ca(OH)$_2$ heterostructure in Sec.~IV.  
In Sec.~V, we show that the optical spectra of the heterostructure can be used to characterize their stacking type. 
Finally, we conclude our results in Sec.~VI.

\section{Computational Methodology}
First-principle calculations are performed using the frozen-core projector augmented wave (PAW)~\cite{paw1} potentials as implemented in the Vienna Ab-initio Simulation Package  (VASP).\cite{vasp1} The electronic exchange-correlation potential is treated within the generalized gradient approximation (GGA) of Perdew-Burke-Ernzerhof (PBE).\cite{pbe1} A plane-wave basis set with kinetic energy cutoff of 500 eV is used. A vacuum spacing of more  than 12~\AA{} is taken to prevent interaction between adjacent images. A set of 20$\times$20$\times$1 $\Gamma$ centered \textbf{k}-point sampling is used for the primitive unit cells. The structures are relaxed until self-consistency for ionic relaxation reached 10$^{-5}$~eV between two consecutive steps. Pressures on the lattice unit cell are decreased to values less than 1.0 kBar in all directions. Atomic charge transfers are calculated using Bader's charge analysis. \cite{bader1} We used the DFT-D2 method of Grimme as implemented in VASP for the
van der Waals correction in all the calculations. \cite{grimme}

The dielectric function and the optical oscillator strength of the individual monolayers and the heterostructure 
are calculated by solving the Bethe-Salpeter equation (BSE) on top of the single shot GW (G$_{0}$W$_{0}$) calculation which is performed on top of standard DFT calculations including 
spin-orbit coupling (SOC). During this process we used $6\times6\times1$ $\Gamma$ centered \textbf{k}-point sampling. The cutoff for the response function was set to 200~eV. The number of bands used in our calculation is 320. The cutoff energy for the plane-waves was chosen to be 400~eV. We include 4 valence and 4 conduction bands into the calculations in the BSE step. 

Once the dielectric function is obtained other linear optical spectral quantities such as layer dependent optical absorbance ($A(\omega)$), transmittance ($T(\omega)$) and frequency dependent reflectivity at normal incidence ($R(\omega)$) can be calculated using the formula;

\begin{equation}
\begin{split}
A(\omega)&=\frac{\omega}{c} L \text{Im}\varepsilon(\omega), \\
T(\omega)&=1-A(\omega),\\
R(\omega)&=\left|\frac{\sqrt{\varepsilon(\omega)}-1}{\sqrt{\varepsilon(\omega)}+1}\right|^2
\end{split}
\end{equation}
where c, L, $\omega$ and $\varepsilon(\omega)$ are the speed of light, the length of the cell in layer-normal direction, frequency of light and 
complex dielectric function, respectively.

\section{G\lowercase{a}S and C\lowercase{a}(OH)$_2$ monolayers}

\begin{figure}[htbp]
\includegraphics[width=13cm]{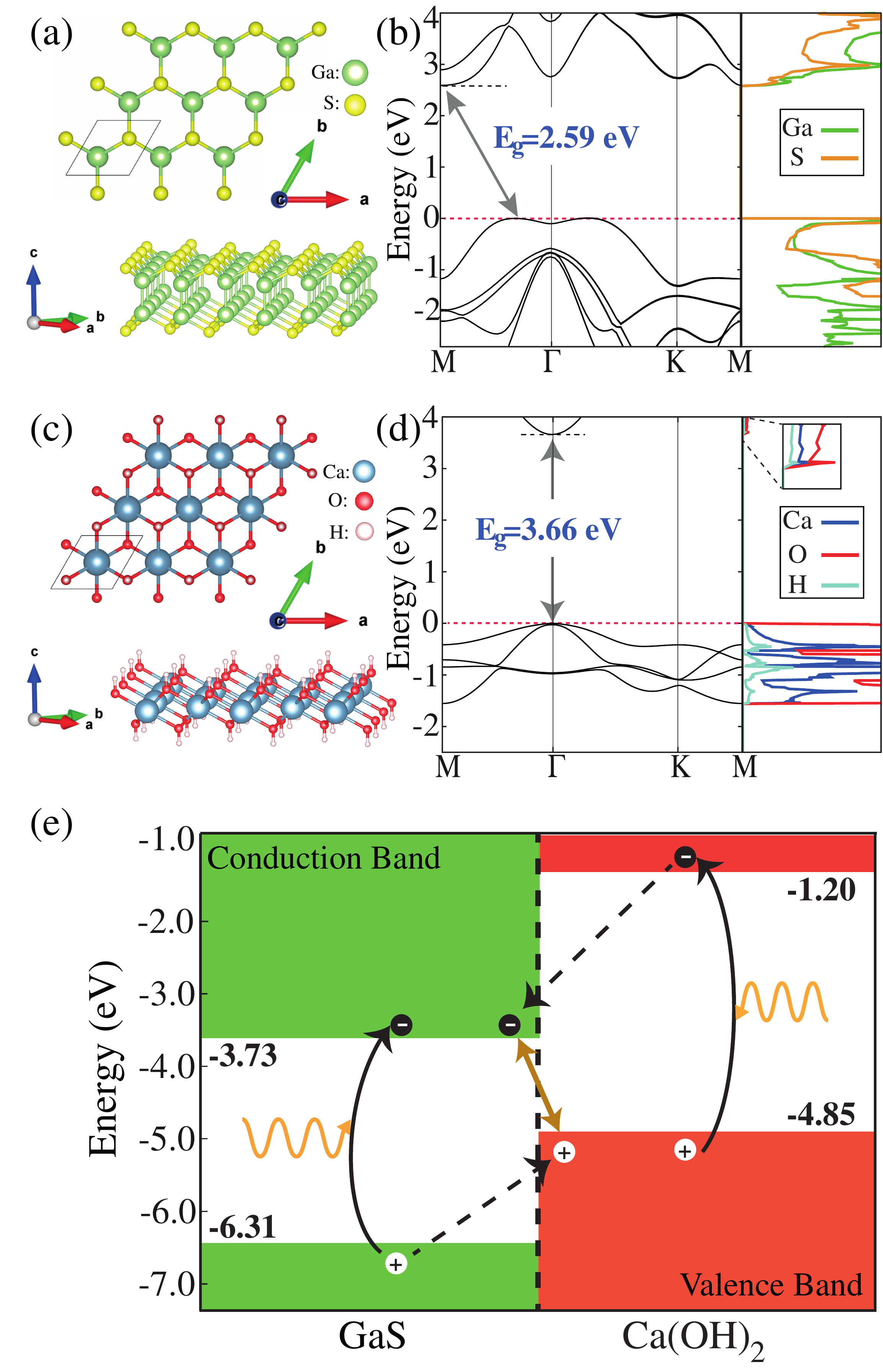}
\caption{\label{fig1} The optimized atomic structures of (a) GaS and (c) 
Ca(OH)$_2$ monolayers and their electronic structures and PDOS (right panel) 
with SOC (b) and (d), respectively. 
The gallium, sulfur, calcium, oxygen and hydrogen atoms are shown 
in green, yellow, blue, red and white, respectively. (e) The calculated band 
alignment of the monolayers where
the vacuum level of both materials is set to 0 eV.}
\end{figure}

The optimized atomic structures of single layers of GaS and Ca(OH)$_2$ are shown 
in Fig.~\ref{fig1}. The optimized lattice parameters for isolated monolayers 
are almost equal to each other, 3.58~\AA~and 3.59~\AA~for GaS and Ca(OH)$_2$, 
respectively. In the GaS monolayer, every Ga atom is covalently bonded to three S 
and one Ga atom which creates a trigonal prismatic symmetry. The distance 
between two nearest neighbor Ga atoms is 2.44~\AA~and the Ga-S distance 
2.36~\AA~is slightly shorter. In the Ca(OH)$_2$ monolayer, the Ca atom has 6 
ionic bonds with O atoms with a bond length of 2.36~\AA. Each O atom in the 
primitive cell has one bond with a  H atom with bond length 0.96~\AA.

The band structures including SOC of the individual monolayers are shown in 
Fig.~\ref{fig1}. Both electronic structures are basically similar to the ones 
reported earlier. \cite{port,ma} The GaS monolayer is an indirect band gap 
semiconductor with a gap of 2.59 eV, as seen in Fig.~\ref{fig1}(b). The valence 
band minimum (VBM) of the compound resides along the $M-\Gamma$ 
direction but the conduction band minimum (CBM) is at the $M$ point in the 
Brillouin zone (BZ). The states in the vicinity of the Fermi level are composed 
of $p$ orbitals, however the VBM is mostly made up of the $s$ orbitals of the Ga and 
S atoms. Our Bader analysis showed that the Ga-S bonds have mostly a covalent 
character and the Ga atoms donate 0.7$e$ whereas each S atom 
gains 0.7$e$ charge.

Contrary to the GaS monolayer, the Ca(OH)$_2$ monolayer is a direct band gap 
semiconductor with a band gap of 3.66 eV. The VBM and CBM of the compound are 
located at the $\Gamma$ point in the BZ. The states at the vicinity of the Fermi 
level originate from the $p_{x}$ and $p_{y}$ orbitals of the O atom, however the 
VBM is mostly from the $s$ and $p_{z}$ orbitals of the Ca and O atoms. The Ca 
atom and H atoms donate, 1.6$e$ and 0.6$e$ charge respectively, and each O atom 
receive 1.4$e$. The bonds in Ca(OH)$_2$ have mostly an ionic character.

Our calculation revealed that the vacuum level of the isolated monolayers are different from each other. 
When the vacuum level of the monolayers is set to 0 eV as  shown in Fig.~\ref{fig1}(e), the obtained heterostructure is a type II heterojunction which opens up the possibility of using
 it as an electron-hole separator under photo-excitation. As shown in the figure, the excited electrons and the holes pile up in the GaS and Ca(OH)$_2$ monolayer, respectively. Since the electrons 
and the holes of the heterostructure reside in different layers, the created excitons are spatially indirect and the recombination occurs through the staggered gap of the heterojuction.
It has been shown that spatially indirect excitons in MX$_2$ heterobilayers have a long lifetime ($\sim$ 20 - 30 ns at room temperature) which is important for applications in 
optoelectronics and photovoltaics. \cite{palummo} 

\begin{figure}[htbp]
\includegraphics[width=13cm]{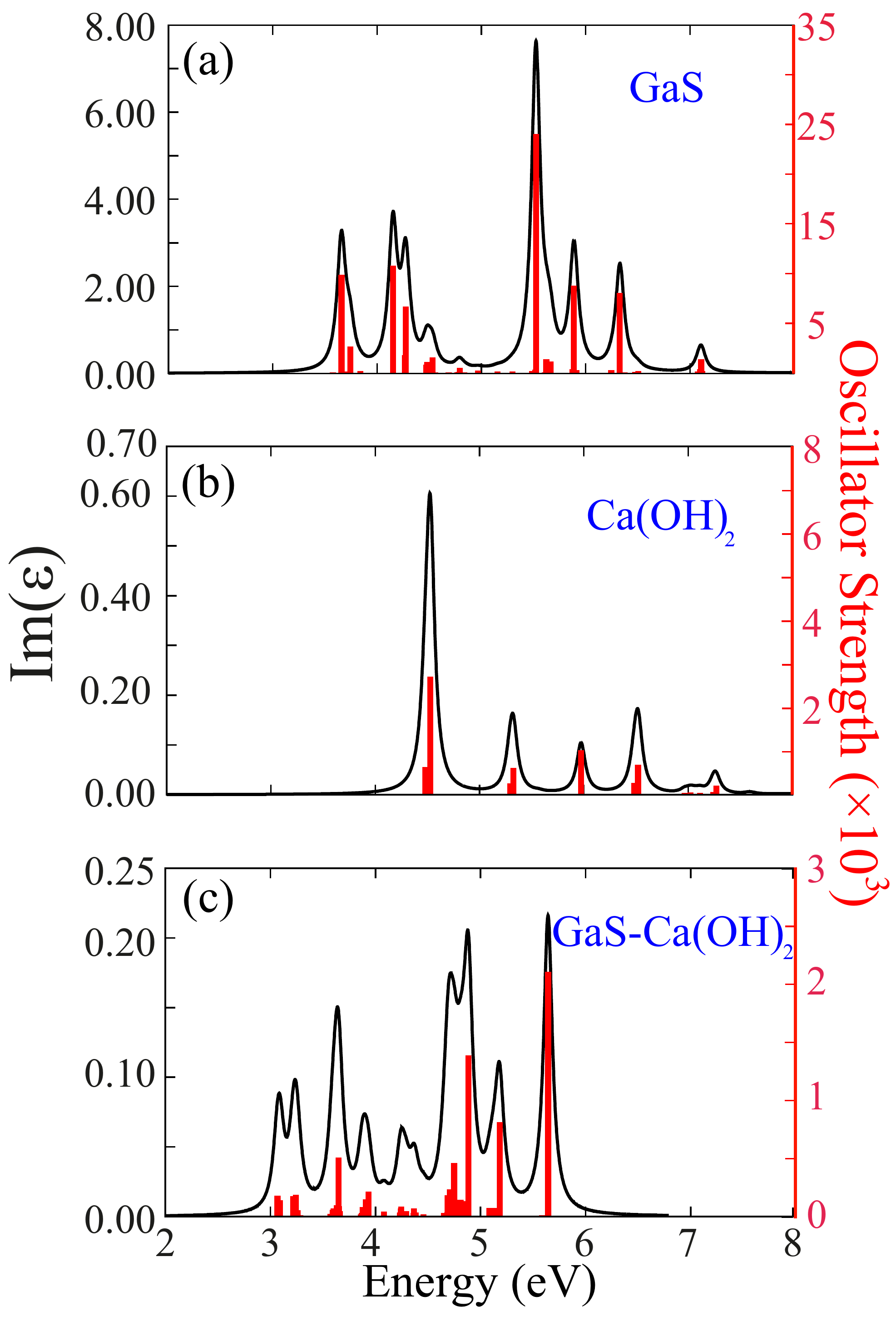}
\caption{\label{fig2} Imaginary part of the dielectric function and the 
oscillator strength of the optical transitions of (a) GaS monolayer (b) 
Ca(OH)$_2$ monolayer and (c) GaS-Ca(OH)$_2$ heterostructure. }
\end{figure}

In Figs.~\ref{fig2}(a) and (b) the dielectric function and the oscillator strength of the optical transitions of isolated GaS and Ca(OH)$_2$ monolayers are shown, respectively.  
For GaS, the first peak of the dielectric function is composed of four optical transitions which are from valence band edge to conduction band edges at M, $\Gamma$ and  
K points in the Brillouin zone (BZ) which are very close in energy. For the Ca(OH)$_2$ monolayer, the first peak is composed of 2 optical transitions from $\Gamma$ to $\Gamma$  in the BZ. These 
two peaks are split by about 40 meV due to the SOC. 
The first peak of the dielectric function is considered as the optical band gap of the compound which is at $\sim$ 3.60 eV and $\sim$ 4.48 eV for GaS and Ca(OH)$_2$, respectively with  exciton binding energy of 1.12 eV and 2.10 eV, respectively. 

\section{Heterostructure} 

\begin{figure}[htbp]
\includegraphics[width=13cm]{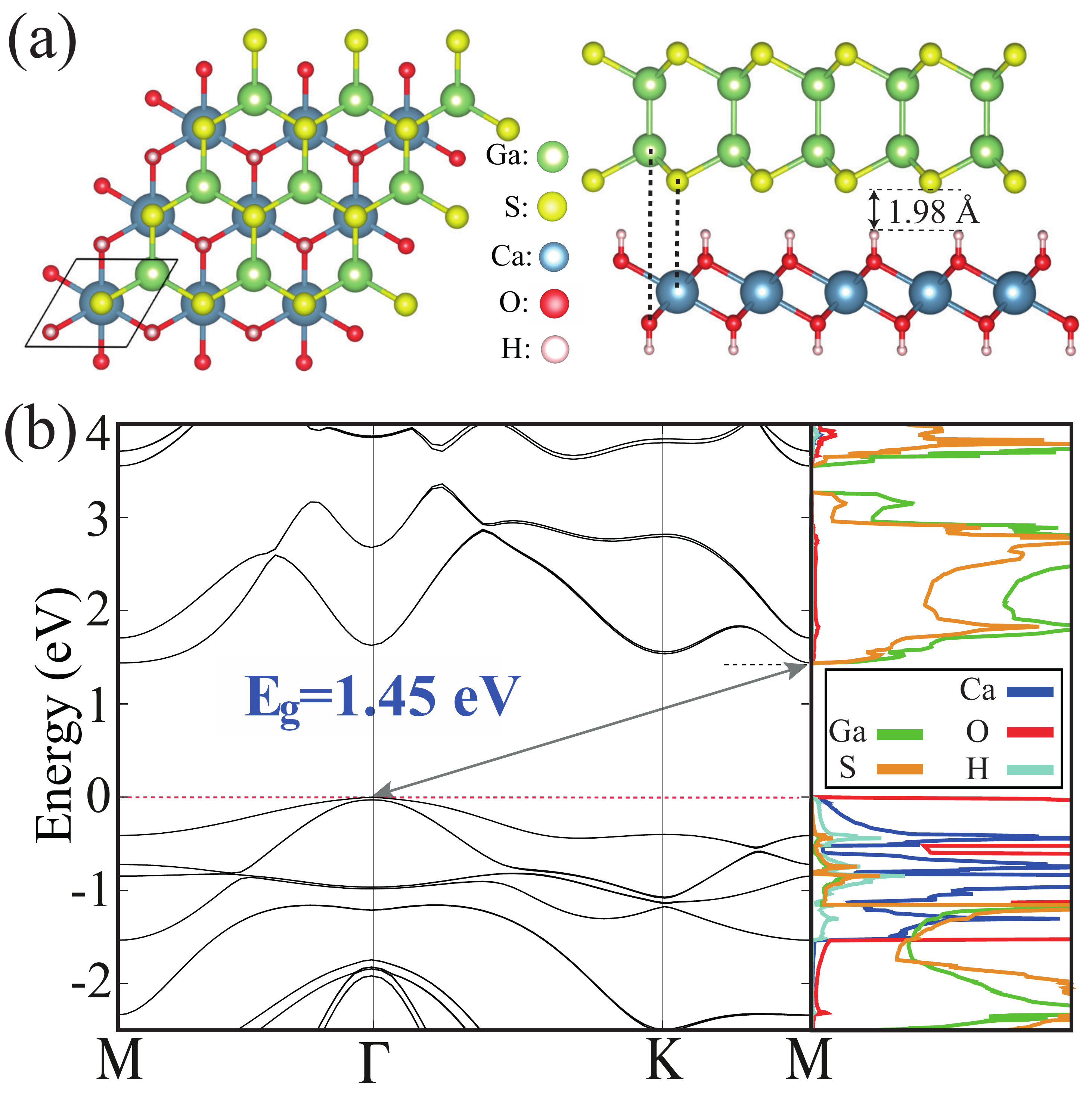}
\caption{\label{fig3} (a) The optimized atomic structures of GaS-Ca(OH)$_2$ 
heterostructure and (b) its electronic band structures together with the PDOS (right panel) 
with SOC. The gallium, sulfur, calcium, oxygen and hydrogen atoms are shown 
in green, yellow, blue, red and white, respectively.}
\end{figure}

Having similar lattice constants opens up the possibility of using these monolayers for building blocks of vdW heterostructures. 
For this purpose we place GaS monolayer on top of Ca(OH)$_2$ monolayer. 
In order to find the minimum energy configuration of the heterostructure we 
shift the GaS monolayer along three different directions. The 
minimum energy configuration is obtained when the Ga and S atoms are 
on top of the O (and H) and Ca atom of the Ca(OH)$_2$ monolayer, 
respectively, as shown in Fig.~\ref{fig3}. The geometric structure of the constituent layers of the heterostructure 
does not change when compared with their isolated form. The distance between the two layers is 1.98~\AA~and the binding 
energy ($E_{B}$) of the heterostructure is 0.12 eV per unit cell (Table.~\ref{table}).

Although the interaction between these two layers are weak, there is a 
significant decrease in the band gap when these two single layers are stacked on 
top of each other (see in Fig.~\ref{fig3}(b)). This dramatic 
modification in the band gap is due to the difference in vacuum level of the 
constituent monolayers. Our calculations show that the heterostructure has an 
indirect band gap of 1.45 eV where the VBM is at the $\Gamma$ and the CBM is at 
the $M$ point in the BZ. The $\Gamma$ to $\Gamma$ gap is 1.63 eV. The advantage 
of having an indirect band gap heterostructure is that it results in long-lived 
excitons due to the small overlap of electron-hole wavefunctions. This opens up 
the possibility of using the GaS-Ca(OH)$_2$ heterojuction for excitonic solar 
cells so that the electron-hole pairs can be split relatively easy.

As shown in Table~\ref{table}, the calculated workfunction of the 
heterostructure is found as 6.21 eV and 5.06 eV for GaS and Ca(OH)$_2$ sides of the heterojunction. 
Due to the interaction between two layers the workfunction values diminished 0.1 and increased 0.21 eV compared with the values for the 
isolated GaS and Ca(OH)$_2$ monolayers, respectively. 

The partial density of states (PDOS) of the GaS-Ca(OH)$_2$ heterostructure is 
shown in the right panel of Fig.~\ref{fig3}(b). As expected from the band 
alignment the valence band and the conduction band of the heterostructure are 
from GaS and Ca(OH)$_2$ monolayers, respectively. The resulting band structure 
of the heterostructure is shown in Fig.~\ref{fig3}(b). As seen the band 
structure of the heterostructure is almost an overlap of the band structures of 
the isolated monolayers. This is due to the weak interaction between layers so 
that the dispersion of the bands does not change.

In order to investigate only the spatially indirect excitons in the 
heterostructure, 4 valence and 4 conduction bands were taken into account for 
the optical transitions in the BSE step. So, the optical transitions between 
these bands correspond to the interlayer recombination of the electrons and 
holes through the staggered gap. This means that the exciton peaks shown in 
Fig.~\ref{fig2}(c) correspond only to spatially indirect excitons in the 
heterojunction. Consistent with the prediction of gap closing in the 
heterostructure, the first peak in the dielectric function appears at lower 
energy ($\sim$ 3.10 eV) than the one of the constituent monolayers as seen in 
the figure. The exciton binding energy of the heterostructure is calculated as 
0.90 eV. 

\begin{table}[htbp]
\caption{\label{table} Ground state properties of GaS and Ca(OH)$_2$ monolayers 
and the heterostructure composed of these two monolayers. 
Calculated lattice parameters $a$ and $b$, interlayer binding energy $E_{B}$ 
(per unit cell), the band gap $E_g$  and the workfunction $\Phi$ (for the heterstructure the first value is for the GaS and the second one is for the Ca(OH)$_2$ side, respectively) of the 
structures.}
\begin{tabular}{ccccccccccccccccc}
\hline\hline
                 & $a$=$b$         &$E_{B}$           &$E_{coh}$/atom   & $E_g$ 
 
  & $\Phi$   &\\
                 &(\AA{}{})         &   (eV)           &   (eV)          &  
(eV) 
   &  (eV)    &\\
\hline
GaS              &    3.58          &   ---           &    3.83         &  
2.59(i) &  6.31    &\\
Ca(OH)$_2$       &    3.59          &   ---           &    4.59         &  
3.66(d) &  4.85    &\\
GaS-Ca(OH)$_2$   &    3.58          &   0.12          &    4.26         &  
1.45(i) &  6.21,5.06    &\\

\hline\hline
\end{tabular}
\end{table}

Our calculations also reveal that the oscillator strengths of 
the GaS monolayer is an order of magnitude larger than that of the Ca(OH)$_2$ 
monolayer and the GaS-Ca(OH)$_2$ heterostructure. However, the oscillator 
strength and the dielectric function of the heterostructure resemble the 
properties of Ca(OH)$_2$. Similarly, as shown by Fang \textit{et 
al.}\cite{fang}, in case of the MoS$_2$-WSe$_2$ heterostructure, which is also a type II 
heterojunction, its optical properties resembles the one of MoS$_2$. Creation of such 
nanoscale type II heterojuctions leads to formation of spatially indirect 
excitons which are vital elements for exciton solar cells and optoelectronics 
devices due to the relatively longer life times of the excitons.

\section{Optical Characterization of Stacking Type}

\begin{figure}
\includegraphics[width=13cm]{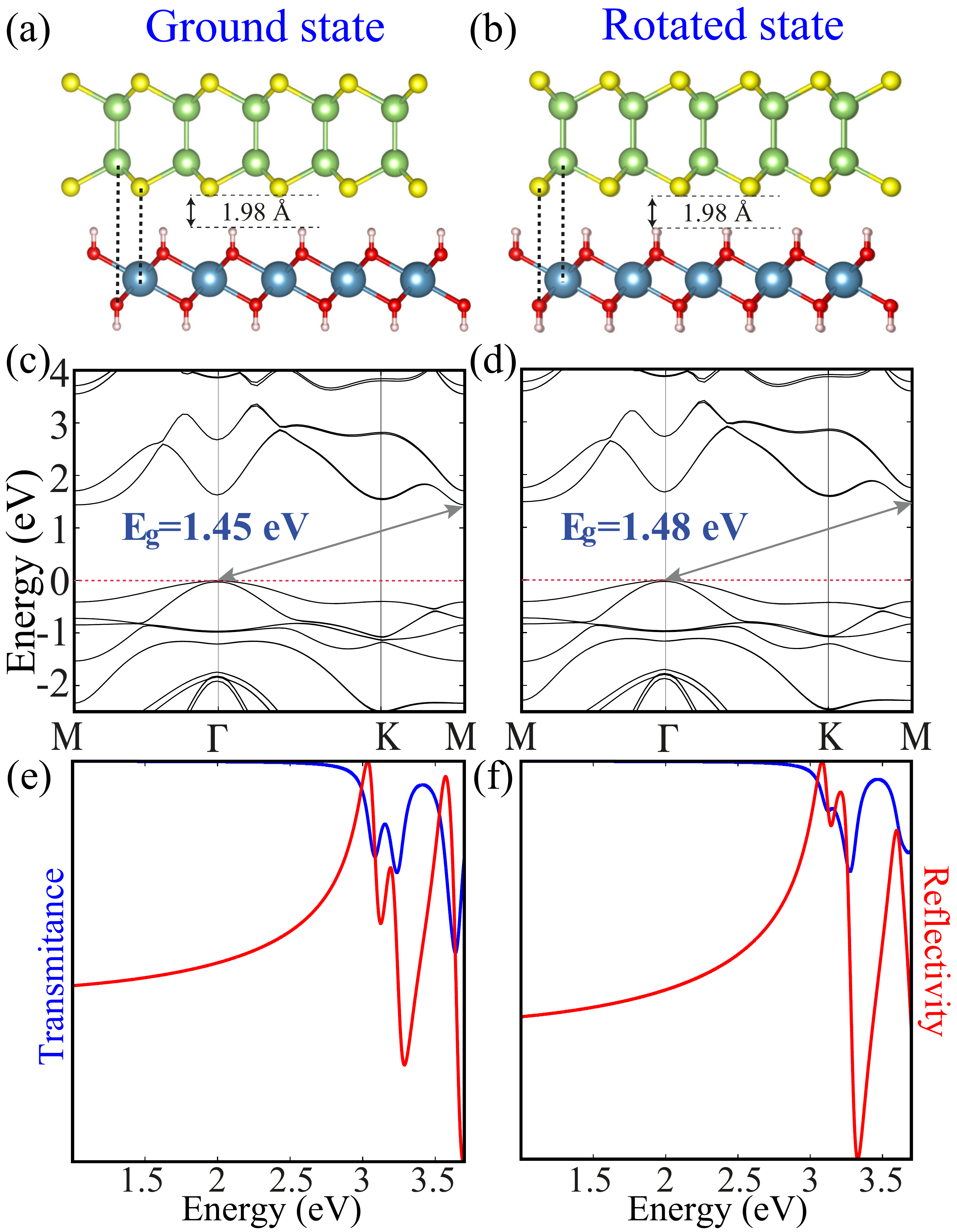}
\caption{\label{fig4} The optimized atomic structures of the two lowest energy 
configurations of GaS-Ca(OH)$_2$ heterostructure; (a) ground and (b) rotated state which has 5 meV higher energy. Their electronic structure ((c) and (d)) and 
optical transmittance and reflectivity ((e) and (f)).}
\end{figure}

In this section, we will show that the optical spectra of ultra-thin heterostructures can be used to determine the stacking type of them.
In Fig.~\ref{fig4}, the optimized atomic configuration, the electronic 
structure and the optical transmittance and the reflectivity of the two lowest 
energy configurations of the GaS-Ca(OH)$_2$ heterostructure are shown. The atomic structures of these two lowest energy configurations are completely different from each other, although the 
energy difference between them is quite small $\sim$ 5 meV.

The higher energy state (Fig.~\ref{fig4}(b)) is obtained by 180$^{\circ}$ rotation of the ground 
state configuration (Fig.~\ref{fig4}(a)), so it will be called as ``rotated state'' from now on. 
In the rotated state the Ga and S atoms are on top of the Ca and O (and H)
atom of the Ca(OH)$_2$ monolayer, respectively and in both cases the interlayer spacing is found as 1.98~\AA. As shown in Figs.~\ref{fig4}(c) and (d), the electronic structure of the ground and 
the rotated state are similar and the dispersion of the bands are identical. The only difference is that the band gap of the rotated state is 3 meV larger.

Although, the electronic structure of the ground and rotated states are  
very close to each other, optical transmittance and the 
reflectivity profiles of their band edges are distinct as shown in Figs.~\ref{fig4}(e) and 
(f). The main differences in the 
optical spectra of these two stacking cases originate from the different 
strength of the optical transitions. It is predicted that the energies of these optical transitions are similar but their intensities are distinct.
This is due to the different orientation of the atoms in these two stackings. 
For instance, the two peaks derived from the band edges are unique for both structures. 
As seen in the optical transmittance profile of the ground state (Fig.~\ref{fig4}(e)) these two peaks can be easily identified because of their almost equal oscillator strength. 
In the rotated case however, the oscillator strength of the first peak is much smaller than the second one so that it can not be identified in the transmittance profile.  
On the other hand, the relative intensity of the first two peaks in the reflectivity profile are also unique for each stacking case. The intensity of the second peak is comparable but 
much smaller as compared to the first peak in the spectrum of the ground state.

This prediction is rather crucial to determine the stacking types of the heterostructures in experiments. The heterojunctions which are produced by growth technique might contain more than one 
stacking types in it, if the energy difference between different stacking types is small. Therefore, it is important to find a powerfull and sensitive tool to identify different stacking types in the produced heterostructures. In the light of above discussion, we suggest that the optical spectra (i.e. transmittance and reflectivity) can be used to identify different stacking types in a heterostructure.

\section{Conclusions}
We proposed a new kind of heterostructure in which the constituent monolayers are GaS and Ca(OH)$_2$ which are recently synthesized. 
These two monolayers have similar lattice constants and they are 
wide band gap semiconductors. When they are stacked on top of 
each other the electronic band gap of the obtained heterostructure decreases 
significantly. Our calculations revealed that the obtained heterojuction is a 
type II semiconductor in which the conduction band and the valence band are from 
GaS and Ca(OH)$_2$, respectively. This leads to localization of the electrons 
and holes in different layers which results in spatially indirect excitons. 
Our BSE-G$_0$W$_0$ calculations showed that the optical 
oscillator strength of the GaS monolayer is an order of magnitude larger than 
the one of monolayer Ca(OH)$_2$. We found that the band edge profiles of the optical spectrum of different stacking 
types of the heterojunction show differences although their electronic structures are rather similar.
This prediction opens up the possibility of using the optical spectrum 
for the characterization of the stacking type of the ultra-thin 
heterostructures. 
 
\section{Acknowledgments} 
This work was supported by the Flemish Science Foundation (FWO-Vl) and the 
Methusalem foundation of the Flemish government. Computational resources were 
provided by TUBITAK ULAKBIM, High Performance and Grid Computing Center 
(TR-Grid e-Infrastructure), and HPC infrastructure of the University of Antwerp 
(CalcUA) a division of the Flemish Supercomputer Center (VSC), which is funded 
by the Hercules foundation. H.S. is supported by a FWO Pegasus long Marie Curie 
Fellowship.


\begin{thebibliography}{99}

\bibitem{graphene1} K. S. Novoselov, A. K. Geim, S. V. Morozov, D. Jiang, Y. Zhang, S. V. Dubonos, I. V. Grigorieva, and A. A. Firsov, Science \textbf{306}, 666 (2004).

\bibitem{graphene2} K. S. Novoselov, D. Jiang, F. Schedin, T. J. Booth, V. V. Khotkevich, S. V. Morozov, and A. K. Geim, Proc. Natl. Acad. Sci. USA \textbf{102}, 10451 (2005).

\bibitem{graphene3} A. K. Geim and K. S. Novoselov, Nat. Mater. \textbf{6}, 183 (2007).

\bibitem{guzman} G. G. Guzm\`{a}n-Verri and L. C. Lew Yan Voon,  Phys. Rev. B \textbf{76}, 075131 (2007).

\bibitem{seymur} S. Cahangirov, M. Topsakal, E. Akturk, H. Sahin, and S. Ciraci, Phys. Rev. Lett. \textbf{102}, 236804 (2009).
 
\bibitem{houssa} M. Houssa, G. Pourtois, V. V. Afanas\'{e}v,  and A. Stesmans, Appl. Phys. Lett. \textbf{96}, 082111 (2010).

\bibitem{sahin} H. Sahin, S. Cahangirov, M. Topsakal, E. Bekaroglu, E. Akturk, R. T. Senger, and S. Ciraci,  Phys. Rev. B \textbf{80}, 155453 (2009).

\bibitem{neto} A. H. C. Neto and K. Novoselov, Rep. Prog. Phys. \textbf{74}, 082501 (2011). 

\bibitem{mak} K. F. Mak, C. Lee, J. Hone, J. Shan, and T. F. Heinz, Phys. Rev. Lett. \textbf{105}, 136805 (2010).

\bibitem{splendiani} A. Splendiani, L. Sun, Y. Zhang, T. Li, J. Kim, C.-Y. Chim, G. Galli, and F. Wang, Nano Lett. \textbf{10}, 1271 (2010).

\bibitem{wang} Q. H. Wang, K. Kalantar-Zadeh, A. Kis, J. N. Coleman, and M. S. Strano, Nat. Nanotechnol. \textbf{7}, 699 (2012). 

\bibitem{port} Y. Aierken, H. Sahin, F. Iyikanat, S. Horzum, A. Suslu, B. Chen, R. T. Senger, S. Tongay, and F. M. Peeters, Phys. Rev. B \textbf{91}, 245413 (2015).

\bibitem{ma} Y. Ma, Y. Dai, M. Guo, L. Yu, and B. Huang, Phys. Chem. Chem. Phys. \textbf{15}, 7098 (2013).

\bibitem{ptmcs} X. Meng, A. Pant, H. Cai, J. Kang, H. Sahin, B. Chen, K. Wu, S. Yang, A. Suslu, F. M. Peeters, and S. Tongay, Nanoscale \textbf{7}, 17109 (2015).

\bibitem{ptmcs2} T. Cao, Z. Li, and S. G. Louie, Phys. Rev. Lett. \textbf{114}, 236602 (2015).

\bibitem{ataca}C. Ataca, H. Sahin, and S. Ciraci, J. Phys. Chem. C \textbf{116}, 8983 (2012).

\bibitem{miro} P. Mir\'{o}, M. Audiffred, and T. Heine, Chem. Soc. Rev. \textbf{43}, 6537 (2014).

\bibitem{chhowalla} M. Chhowalla, H. S. Shin, H. S. Shin, L.-J. Li, K. P. Loh, and H. Zhang, Nat. Chemistry \textbf{5}, 263 (2013).

\bibitem{he} K. He, N. Kumar, L. Zhao, Z. Wang, K. F. Mak, H. Zhao, and J. Shan, Phys. Rev. Lett. \textbf{113}, 026803 (2014). 

\bibitem{klots} A. R. Klots, A. K. M. Newaz, B. Wang, D. Prasai, H. Krzyzanowska, J. Lin, D. Caudel, N. J. Ghimire, J. Yan, B. L. Ivanov, K. A. Velizhanin, A. Burger, D. G. Mandrus, N. H. Tolk, 
S. T. Pantelides, and  K. I. Bolotin, Scientific Reports \textbf{4}, 6608 (2014).

\bibitem{chernikov} A. Chernikov, T. C. Berkelbach, H. M. Hill, A. Rigosi, Y. Li, O. B. Aslan, D. R. Reichman, M. S. Hybertsen, and T. F. Heinz, Phys. Rev. Lett. \textbf{113}, 076802 (2014). 

\bibitem{ugeda} M. M. Ugeda, A. J. Bradley, S.-F. Shi, F. H. da Jornada, Y. Zhang, D. Y. Qiu, W. Ruan, S.-K. Mo, Z. Hussain, Z.-X. Shen, F. Wang, Steven G. Louie, and Michael F. Crommie, 
Nat. Mat. \textbf{13}, 1091 (2014).

\bibitem{heinz} K. F. Mak, K. He, J. Shan, and Tony F. Heinz,  Nat. Nanotechnol. \textbf{7}, 494 (2012). 

\bibitem{sallen} G. Sallen, L. Bouet, X. Marie, G. Wang, C. R. Zhu, W. P. Han, Y. Lu, P. H. Tan, T. Amand, B. L. Liu, and B. Urbaszek, Phys. Rev. B \textbf{86}, 081301(R) (2012).

\bibitem{xu} X. Xu, W. Yao, D. Xiao, and T. F. Heinz, Nat. Phys. \textbf{10}, 343 (2014). 

\bibitem{late1} D. J. Late, B. Liu, H. S. S. R. Matte, C. N. R. Rao, and V. P. Dravid, Adv. Funct. Mater. \textbf{22}, 1894 (2012).

\bibitem{late2} D. J. Late, B. Liu, J. Luo, A. Yan, H. S. S. R. Matte, M. Grayson, C. N. R. Rao, and V. P. Dravid, Adv. Funct. Mater. \textbf{24}, 3549 (2012).

\bibitem{hu}  P. A. Hu, Z. Z. Wen, L. F. Wang, P. H. Tan, and K. Xiao, ACS Nano \textbf{6}, 5988 (2012).

\bibitem{geim} A. K. Geim and I. V. Grigorieva, Nature (London) \textbf{499}, 419 (2013). 

\bibitem{ross} J. S. Ross, P. Klement, A. M. Jones, N. J. Ghimire, J. Yan, D. G. Mandrus, T. Taniguchi, K. Watanabe, K. Kitamura, W. Yao, D. H. Cobden, and X. Xu, Nat. Nanotechnol. \textbf{79}, 
268 (2014).

\bibitem{furchi} M. M. Furchi, A. Pospischil, F. Libisch, J. Burgd\"{o}rfer, and T. Mueller, Nano Lett. \textbf{14}, 4785 (2014).

\bibitem{roy} T. Roy, M. Tosun, X. Cao, H. Fang, D.-H. Lien, P. Zhao, Y.-Z. Chen, Y.-L. Chueh, J. Guo, and A. Javey, ACS Nano \textbf{9}, 2071 (2015).

\bibitem{rivera} P. Rivera, J. R. Schaibley, A. M. Jones, J. S. Ross, S. Wu, G. Aivazian, P. Klement, K. Seyler, G. Clark, N. J. Ghimire, J. Yan, D. G. Mandrus, W. Yao, and X. Xu, Nat. Commun. 
\textbf{6}, 6242 (2015).

\bibitem{fang} H. Fang, C. Battaglia, C. Carraro, S. Nemsak, B. Ozdol, J. S. Kang, H. A. Bechtel, S. B. Desai, F. Kronast, A. A. Unal, G. Conti, C. Conlon, G. K. Palsson, M. C. Martin, Andrew
 M. Minor, C. S. Fadley, E. Yablonovitch, R. Maboudian, and A. Javey, Proc. Natl. Acad. Sci. USA \textbf{111}, 6198 (2014).

\bibitem{calman} E. V. Calman, C. J. Dorow, M. M. Fogler, L. V. Butov, S. Hu, A. Mishchenko, and A. K. Geim,  arXiv:1510.04410v1.

\bibitem{paw1} G. Kresse and D. Joubert, Phys. Rev. B \textbf{59}, 1758 (1999).

\bibitem{vasp1} G. Kresse and J. Hafner, Phys. Rev. B \textbf{47}, 558(R) (1993).

\bibitem{pbe1} J. P. Perdew, K. Burke, and M. Ernzerhof, Phys. Rev. Lett. \textbf{77}, 3865 (1996).

\bibitem{bader1} R. F. W. Bader, \textit{Atoms in Molecules - A Quantum Theory} (Oxford University Press, Oxford, UK, 1990).

\bibitem{grimme} S. Grimme, J. Comput. Chem. \textbf{27}, 1787 (2006).

\bibitem{palummo} M. Palummo, M. Bernardi, and J. C. Grossman, Nano Lett. \textbf{15}, 2794 (2015). 

\end{thebibliography}
\end{document}